\documentclass[aps,twocolumn]{revtex4}

\usepackage{graphicx}
\usepackage{pslatex}
\usepackage{epstopdf}
\usepackage{amssymb}
\usepackage{mathrsfs}
\usepackage{amsmath}
\usepackage{bbm}
\usepackage[utf8]{inputenc} 
\usepackage{multirow} 
\usepackage{float} 
\usepackage[usenames]{color} 
\usepackage[hidelinks]{hyperref} 
\usepackage{ragged2e} 

\newcommand{\beq}{\begin{equation}}
\newcommand{\eeq}{\end{equation}}

\def\la{{\langle}}
\def\ra{{\rangle}}

\begin{document}

\title{Five  Experimental Tests  on the  5-qubit IBM Quantum Computer}

\vspace{.5cm}

\author{Diego Garc\'{\i}a-Mart\'{\i}n   and Germ\'an Sierra}

\address{Instituto de F\'{\i}sica Te\'orica (IFT) UAM-CSIC, Universidad Autónoma de Madrid, Cantoblanco, Madrid, Spain}

\bigskip\bigskip\bigskip\bigskip

	\begin{abstract} \normalsize
		The 5-qubit quantum computer prototypes that IBM has given open access to on the cloud allow the implementation of real experiments on a quantum processor. We present the results obtained in five experimental tests performed on these computers: dense coding, quantum Fourier transforms, Bell's inequality, Mermin's inequalities (up to $n=5$) and the construction of the prime state $|p_3\ra$. These results serve to assess the functioning of the IBM 5Q chips.
		
		
	\end{abstract}

\maketitle

\vskip 0.2cm

\setlength\parskip{0.1cm}

\section{Introduction}

Quantum Computation has become a very exciting, promising and active field of research for hundreds of scientists around the world during the last decades (see e.g. \cite{Chuang,Merminbook}). After the important theoretical discoveries in the 90s (Shor's algorithm, quantum error correction, quantum computers as universal quantum simulators...), the advances in Experimental Physics and Engineering have made possible to build the first quantum-computer prototypes.


In this respect, IBM released in 2016 a 5-qubit universal quantum computer prototype accessible on the cloud, based on superconducting qubits: the IBM Quantum Experience \cite{IBMQ}. About a year later, IBM enlarged its Quantum Experience with the release under restricted access of a 16-qubit universal quantum computer (also made of superconducting qubits), and announced that they had designed a commercial prototype of a 17-qubit quantum processor. And just a few days ago (November 2017), IBM announced that they have already built a 20-qubit and a 50-qubit quantum computer, with decoherence times that double those exhibited by the computers of the Quantum Experience.




	

The quantum computer prototypes of the IBM Quantum Experience --namely \emph{ibmqx2} (5 qubits), \emph{ibmqx4} (5 qubits) and \emph{ibmqx5} (16 qubits)-- allow the implementation by the scientific community of real experiments on a quantum processor (see \cite{Latorre, Qcheque, Majorana, error, compressed, Cloud, entangled, entropy, repetition, Benchmark, phasespace, noncontextual, D-Jozsa, AE, No-hidding, envariance, permutation, indios1, indios2}). Recent applications include, for instance, the design of a quantum cheque \cite{Qcheque}, a simulation of the braiding of two non-abelian anyons \cite{Majorana} or a demonstration of fault-tolerant quantum computation \cite{error}.

In order to make proper use of these computers, some technical aspects must be taken into account. First of all, the set of gates available includes the Hadamard gate $H$, the $X,Y,Z$ gates (these are the usual Pauli matrices), the $S,T,S^\dagger,T^\dagger$ rotation gates and a parameter-dependent rotation, which introduces a relative phase $e^{i\,\lambda}$ in the state of the qubit. There are another two parameter-dependent transformations available as well, but those will not be used in the present work. And there is a 2-qubit gate: the controlled-NOT. Any desired unitary transformation must be accomplished with just these gates.

Another technical detail is that not all the qubits are connected among themselves due to experimental constraints. This means that controlled-NOT operations (cNOTs) are restricted to some particular pairs of qubits, as shown in Fig.1 (this fact turns out to be relevant in the present implementation of the quantum computer, because it increases the number of gates needed for a given circuit, leading to a decrease in performance).

\begin{center}
	\includegraphics[scale=0.66]{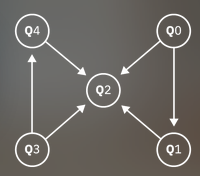}\;\;\;\; \includegraphics[scale=0.66]{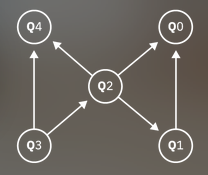}
	
	\justifying \noindent\begin{small}\textbf{Figure 1}. Diagram of the available cNOTs among qubits on the IBM 5Q computers: ibmqx2 (left) and ibmqx4 (right). The qubits are represented by circles, while the cNOTs are arrows pointing from control qubit to target qubit.\end{small}
\end{center}


There exists, however, an important identity that reverses the control and the target qubits of a cNOT using four Hadamard gates (shown in Fig.2 below). This identity allows to use cNOTs in any direction among the qubits connected by arrows in Fig.1.

\begin{center}
	\includegraphics[scale=0.38]{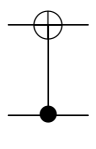}\includegraphics[scale=0.4]{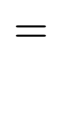} \includegraphics[scale=0.37]{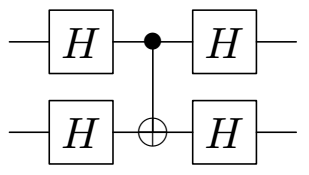}
	
	\justifying\noindent\begin{small}\textbf{Figure 2}. The control and target qubits of a cNOT are reversed by four Hadamard gates.\end{small}
\end{center}

Other identities that will be used are shown in Figs.3-5. The implementation of a Toffoli gate in terms of 1-qubit gates and cNOTs \cite{Chuang} is shown in Fig.6. We shall also use the fact that the square of the Hadamard gate is the identity ($H^2=\mathbbm{1}$), which often leads to simplifications of the final circuits.

\begin{center}
	\includegraphics[scale=0.38]{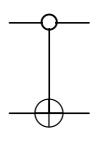} \includegraphics[scale=0.38]{equal} \includegraphics[scale=0.38]{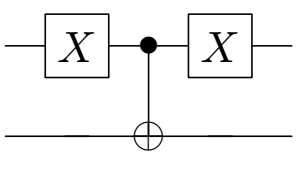}
	
	\justifying\noindent\begin{small}\textbf{Figure 3}. A controlled-NOT that acts on the target qubit when the control qubit is in the state $|0\ra$.\end{small}
\end{center}

\begin{center}
	\includegraphics[scale=0.45]{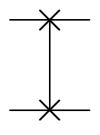} \includegraphics[scale=0.42]{equal} \includegraphics[scale=0.45]{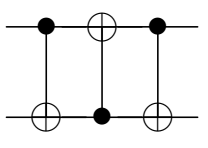}
	
	\justifying\noindent\begin{small}\textbf{Figure 4}. A SWAP between two qubits is equivalent to three consecutive cNOTs with the one in the middle reversed. \end{small}
\end{center}

\begin{center}
	\includegraphics[scale=0.38]{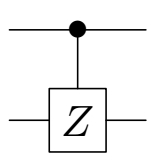} \includegraphics[scale=0.42]{equal} \includegraphics[scale=0.38]{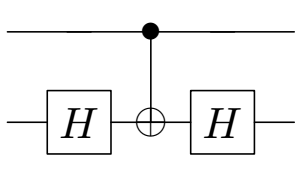}
	
	\justifying\noindent\begin{small}\textbf{Figure 5}. A controlled-$Z$ gate in terms of a cNOT and two Hadamard gates.\end{small}
\end{center}

\begin{center}
	\includegraphics[scale=0.23]{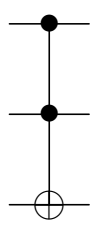} \includegraphics[scale=0.355]{equal} \includegraphics[scale=0.23]{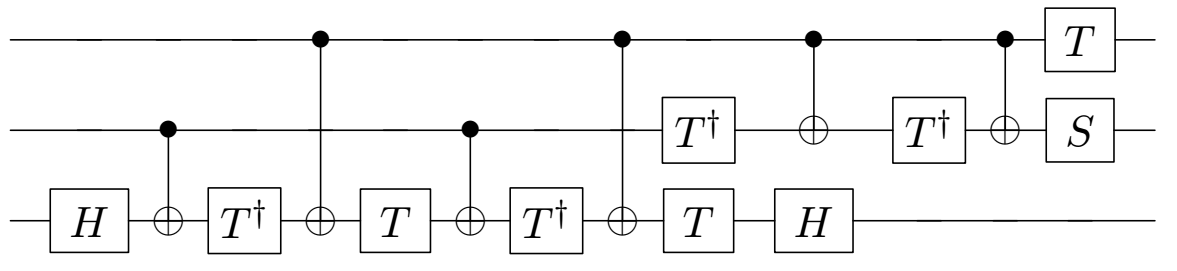}
	
	\justifying\noindent\begin{small}\textbf{Figure 6}. A Toffoli gate implemented as the product of 1-qubit gates and cNOTs.\end{small}
\end{center}

Finally, measurements on the IBM 5Q computers can only be carried out in the computational basis $\{|0\ra,\,|1\ra\}$. However, measurements in other basis can be simulated by means of appropriate gates. In this way, a Hadamard gate prior to measurement in the computational basis is equivalent to a measurement in the $X$ basis $\left\{|+\ra=\frac{|0\ra+|1\ra}{\sqrt{2}}\},\,|-\ra=\frac{|0\ra-|1\ra}{\sqrt{2}}\right\}$, and a $HS^\dagger$ gate is equivalent to a measurement in the $Y$ basis $\left\{|+\ra_y=\frac{|0\ra+i\,|1\ra}{\sqrt{2}}\},\,|-\ra_y=\frac{|0\ra-i\,|1\ra}{\sqrt{2}}\right\}$.

\vspace{0.5cm}

With these basic notions in mind, we have carried out several experiments and protocols: dense coding, quantum Fourier transforms, Bell's inequality, Mermin's inequalities (up to $n=5$) and the construction of the prime state $|p_3\ra$. The IBM Quantum Experience enables the simulation of the circuits prior to its actual implementation on the real quantum computer; this is helpful to ensure that the circuits are well designed. Each circuit has been run 5 times, each run comprising 8192 repetitions or shots. Mean results and standard deviations among the five runs have been calculated in all cases.

\section{Implemented Experiments}

\noindent\textbf{1. Dense coding}

\vspace{0.1cm}

Dense coding is a protocol introduced by Bennett and Wiesner \cite{BW} that allows two bits of classical information to be transmitted between two partners (Alice and Bob) that share an EPR pair by performing local operations on just a single qubit (Alice's) of the entangled pair, which is then eventually sent to Bob.

Dense coding can be implemented on the IBM 5Q using the following circuit, designed by Mermin \cite{DenseC}:

\begin{center}
	\includegraphics[scale=0.31]{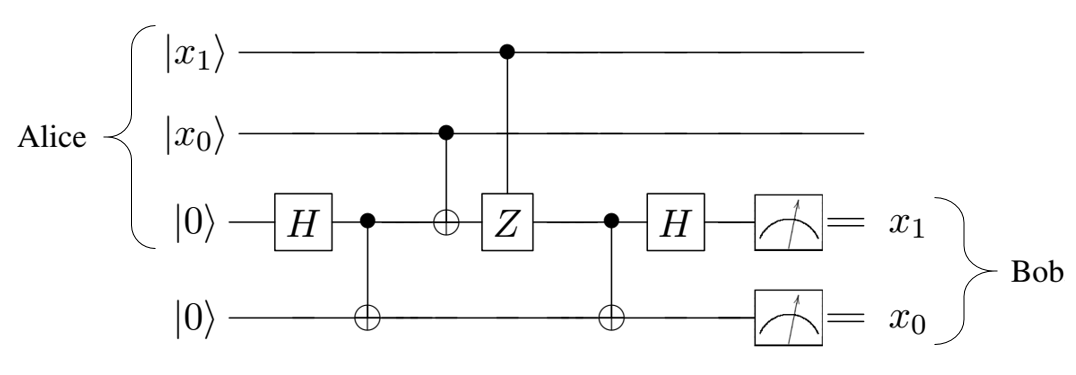}
	
	\justifying\noindent\begin{small}\textbf{Figure 7}. Circuit for the implementation of dense coding.\end{small}
\end{center}

The two uppermost horizontal wires in the circuit in Fig.7 represent the two classical bits that are to be sent by Alice (i.e. $|x_i\ra=\{|0\ra,\,|1\ra\}$); this two bit string is generated using $X$ gates (i.e. $|x_1x_0\ra=X^{x_1}\otimes X^{x_0}\,|00\ra$). The two lowermost horizontal wires represent the qubits shared by Alice and Bob. The initial Hadamard gate and cNOT generate the entangled pair in the state $|\phi^+\ra=\frac{1}{\sqrt{2}}\,(|00\ra+|11\ra)$. Then, the second cNOT and the controlled-Z implement the transformations of the protocol. Finally, the third cNOT and the last Hadamard gate transform the resulting Bell state into one of the four computational basis states before a joint measurement in this basis is carried out by Bob in order to obtain the information sent over by Alice. 


The results obtained using the ibmxq4 are shown in Table 1:

\vspace{0.2cm}

\begin{table}[H]
	\begin{center}
		\begin{tabular}{p{0.75cm}|p{1.73cm}p{1.7cm}p{1.7cm}p{1.7cm}}
			$\;\;A\;\backslash \; B\,$ & \centering 00 			 & \centering 01 	 		  & \centering 10    		   &  \quad \quad11  \\ \hline
			\centering 00 & \textbf{0.826$\pm$0.003} & 0.041$\pm$0.002 & 0.111$\pm$0.004 & 0.023$\pm$0.003 \\ \hline
			\centering01 & 0.114$\pm$0.007 & \textbf{0.744$\pm$0.005} & 0.039$\pm$0.003 & 0.101$\pm$0.006 \\ \hline
			\centering10 & 0.159$\pm$0.025 & 0.025$\pm$0.001	& \textbf{0.778$\pm$0.025} & 0.038$\pm$0.002 \\ \hline
			\centering11 & 0.044$\pm$0.005 & 0.097$\pm$0.010 & 0.114$\pm$0.003  & \textbf{0.746$\pm$0.013} \\ \hline
		\end{tabular}
	\end{center}
		\begin{small}\textbf{Table 1}. Mean probability outcomes ($\pm$ standard deviation) of the dense-coding circuit from Fig.7 after 5 runs of 8192 shots on the IBM 5Q computer (ibmqx4), using qubits 3,2,1,0. The column on the leftmost edge shows the 2-bit string transmitted by Alice and the uppermost row shows the possible outcomes after a joint measurement performed by Bob. In bold are the outcomes expected with probability equal to 1 in each case.\end{small}
\end{table}

The protocol is therefore successfully completed around $83 \%$, $74 \%$, $78 \%$ and $75 \%$ of the times for the sequences $00$, $01$, $10$ and $11$, respectively. 

Certainly, the complete protocol should be carried out in different locations and include the physical transmission of one of the entangled qubits in order to be of practical use, but a \emph{proof of principle} as the one shown here seems to be a necessary previous step before the experimentally much more complicated full protocol can take place (a previous step that, as the results show, it is not completed successfully with sufficiently high accuracy yet). It can be argued that superconducting qubits are not the ideal system to be sent over large distances, but then it should be considered how dense coding is going to be incorporated into a quantum computer that functions with superconducting qubits (if it is going to be so at all).

\vspace{0.3cm}

\noindent\textbf{2. Quantum Fourier transform}

\vspace{0.1cm}

The quantum Fourier transform (QFT) is a basic unitary transformation in the field of Quantum Computation. For a given state of the computational basis $|j\,\ra\equiv|j_{n-1}\cdot\cdot\cdot j_1j_0\ra$, with $j_i=\{0,1\}$, it is defined (in its product representation) as:

\beq \label{F} \begin{split} &\quad|j_{n-1}\cdot\cdot\cdot j_1j_0\ra\;\longrightarrow\\\\\longrightarrow\;\frac{1}{\sqrt{2^n}}\,(|0\ra+&e^{2\pi i\; 0.j_0}\,|1\ra)\,(|0\ra+e^{2\pi i\; 0.j_1j_0}\,|1\ra)\cdot\cdot\cdot\\\\\cdot\cdot\cdot&\,(|0\ra+e^{2\pi i\; 0.j_{n-1}j_{n-2}...j_0}\,|1\ra) \quad ,\end{split} \eeq








\noindent where $0.j_{n-1}j_{n-2}...j_0=j_{n-1}/2+j_{n-2}/2^2+...j_0/2^n$ represents a binary fraction. An efficient way of implementing the QFT for states of three qubits is shown in Fig.8 below:

\begin{center}
	\includegraphics[scale=0.36]{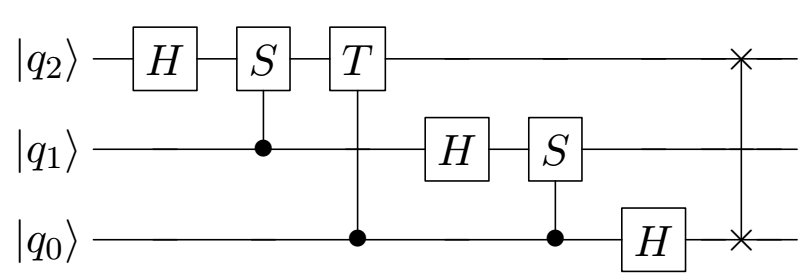}
	
	\justifying\noindent\begin{small}\textbf{Figure 8}. Circuit implementing the quantum Fourier transform of a state of three qubits.\end{small}
\end{center}



	



To implement the circuit shown in Fig.8 on the IBM 5Q, we use the following identity:

\begin{center}
	\includegraphics[scale=0.34]{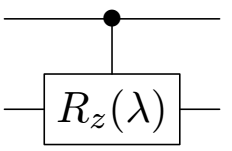} \includegraphics[scale=0.36]{equal}  \includegraphics[scale=0.34]{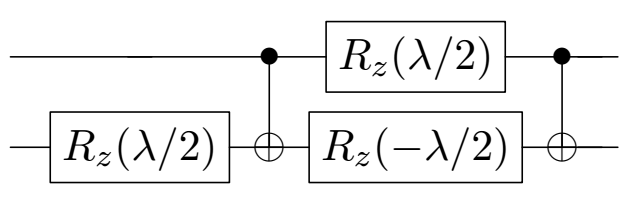} 
	
	\justifying\noindent\begin{small}\textbf{Figure 9}. Implementation of a controlled-$R_z(\lambda)$ gate with 1-qubit gates and cNOTs.\end{small}
\end{center}

\noindent where $R_z(\lambda)$ is a rotation of $\lambda$ radians around the $Z$ axis on the Bloch sphere:

\beq R_z(\lambda)\equiv \left(\begin{matrix}
	1 & 0 \\  0 & e^{\,i\lambda}
\end{matrix}\right) \quad.\eeq

The performance of the QFT of different computational-basis states can be assessed using a simple procedure that takes into account the fact that the state of each qubit in the product state (\ref{F}) lies on the equator of the Bloch sphere. Hence, applying  $R_z(\lambda)$ gates to all qubits (with an appropriate value of $\lambda$ for each qubit), the state of every qubit can be taken to the closest of $\{|+\ra,\,|-\ra\}$ on the sphere. In this way, finally measuring in the $X$ basis should result in a single three-bit string with probability equal to 1 in the ideal case.

For the 3-qubit case, the initial states chosen to be Fourier transformed are $|000\ra$ and $|011\ra$. For the state $|000\ra$, the QFT --equation (\ref{F})-- gives the state:

\beq \label{f1} \frac{1}{2\sqrt{2}}\,(|0\ra+|1\ra)\,(|0\ra+|1\ra)(|0\ra+|1\ra) \eeq

\noindent while the QFT of the state $|011\ra$ gives:

\beq \label{f2} \frac{1}{2\sqrt{2}}\,(|0\ra-|1\ra)\,(|0\ra+e^{\,3\pi i/2}\,|1\ra)(|0\ra+e^{\,3\pi i/4}\,|1\ra) \eeq

The results obtained for the two states (\ref{f1}) and (\ref{f2}) on the ibmqx4 are shown in Table 2:

\vspace{0.2cm}
\begin{table}[H]
	\begin{center}
		\begin{tabular}{c|p{0.75cm}p{0.75cm}p{0.75cm}p{0.75cm}p{0.75cm}p{0.75cm}p{0.75cm}p{0.75cm}}
			& 000 & 001 & 010 & 011 & 100 & 101 & 110 & 111  \\ \hline
			$|000\ra\;$ & \textbf{0.748} & 0.076 & 0.053 & 0.020 & 0.055 & 0.025 & 0.016 & 0.008 \\ \hline
			$|011\ra\;$ & 0.042 & 0.104 & 0.026 & 0.043 & 0.078 & \textbf{0.615} & 0.020 & 0.071 \\ \hline
		\end{tabular}
	\end{center}
	\begin{small}\textbf{Table 2}. Mean probability outcomes of the Fourier transform of the states on the leftmost column after 5 runs of 8192 shots on the IBM 5Q computer (ibmqx4), using qubits 2,1,0 (standard deviations are not shown for the sake of clarity). Measurements were carried out in an appropriate basis so that the expected result is 000 for the state $|000\ra$ and 101 for the state $|011\ra$ (both marked in bold), as explained in the text.\end{small}

\end{table}

Since, as explained above, the expected results are 000 and 101 with probability equal to 1 for the QFT of the states $|000\ra$ and $|011\ra$ respectively, it can be said that these states have been Fourier transformed with reliabilities of around $74.8\pm1.1\%$ and $61.5\pm1.0\%$ (the errors are the standard deviations among the 5 runs). QFTs of states of more than three qubits were not tried out because the limited connectivity among qubits (shown in Fig.1) would result in low-performance circuits.


\vspace{0.4cm}

\noindent\textbf{3. Bell's inequality}

\vspace{0.1cm}

Bell's theorem states that no deterministic local hidden-variable theory can reproduce all predictions of Quantum Mechanics \cite{Bell}. Bell derived an inequality for the singlet-spin state (equivalent to $|\psi^-\ra=\frac{1}{\sqrt{2}}\,(|01\ra-|10\ra)$), known as Bell's inequality, which must be fulfilled for any local realistic (hidden-variable) theory:


\beq \label{Bell} |P(\vec{a},\vec{b})-P(\vec{a},\vec{c})|-P(\vec{b},\vec{c})\leq 1 \quad,\eeq

\noindent where $P(\vec{a},\vec{b})$ is the mean value of the product of the outcomes of measuring the spin components of two entangled spin $1/2$ particles in the state $|\psi^-\ra$ in the directions $\vec{a}$ and $\vec{b}$ respectively (being the possible outcomes $\pm1$), and analogously for $P(\vec{a},\vec{c})$ and $P(\vec{b},\vec{c})$. The quantum-mechanical expectation value for $P(\vec{a},\vec{b})$ is:

\beq \label{QM} P(\vec{a},\vec{b})_{QM}= \la \psi^-|\vec{\sigma}\cdot{\vec{a}}\otimes\vec{\sigma}\cdot{\vec{b}}\,|\psi^-\ra =-cos\,\theta_{\,\vec{a},\,\vec{b}} \quad,\eeq

\noindent where $\theta_{\,\vec{a},\,\vec{b}}\;$ is the angle between $\vec{a}$ and $\vec{b}$. This means that the angles that produce maximal violation of inequality (\ref{Bell}) are $\theta_{\,\vec{a},\,\vec{b}}=\theta_{\,\vec{b},\,\vec{c}}=\pi/3$; that is, angles of $60^{\circ}$ between $\vec{a},\vec{b}$ and $\vec{b},\vec{c}$, and $120^{\circ}$ between $\vec{a},\vec{c}$ (see Fig.10). According to Quantum Mechanics, the inequality should then be violated as $1.5\nleq1$.



\begin{center}
	\includegraphics[scale=0.65]{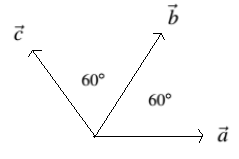}
	
	\justifying\noindent\begin{small}\textbf{Figure 10}. Directions that produce maximal violation of Bell's inequality.\end{small}
\end{center}

Bell's inequality (\ref{Bell}) can be tested on the IBM 5Q computers using the following three circuits:

\begin{center}
	\includegraphics[scale=0.36]{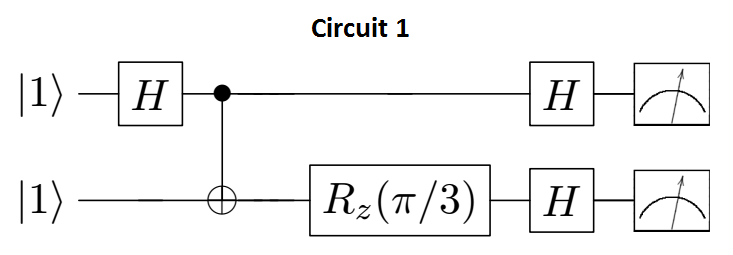}
\end{center}

\begin{center}	
 \includegraphics[scale=0.36]{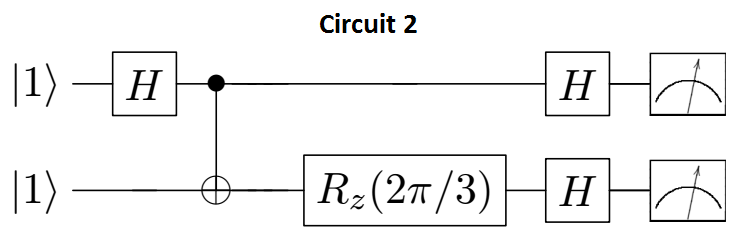}
\end{center}

\begin{center}
	\includegraphics[scale=0.36]{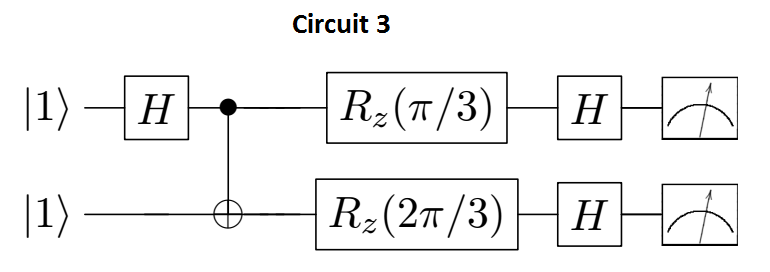}
	
	\justifying\noindent\begin{small}\textbf{Figure 11}. Circuits for measuring 1) $P(\vec{a},\vec{b})$ 2) $P(\vec{a},\vec{c})$ and 3) $P(\vec{b},\vec{c})$.\end{small}
\end{center}

The Hadamard gate and cNOT on the circuits from Fig.11 generate the state $|\psi^-\ra$ from $|11\ra$. Then, the first circuit performs measurements in the $X$ basis ($\vec{a}$) and in a basis whose vectors form a $\pi/3$ angle with those of the $X$ basis ($\vec{b}$), so it can be used to measure $P(\vec{a},\vec{b})$. The second and third circuits measure $P(\vec{a},\vec{c})$ and $P(\vec{b},\vec{c})$, respectively. The results obtained using the ibmqx4 are shown in Fig.12:


\begin{center}
	\includegraphics[scale=0.55]{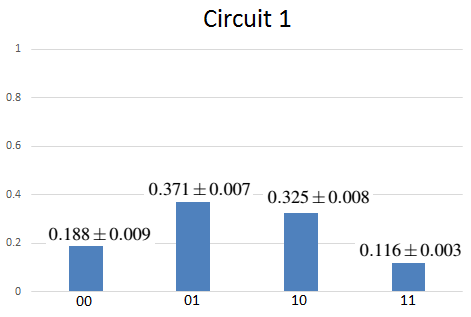}
\end{center}

\begin{center}
 \includegraphics[scale=0.55]{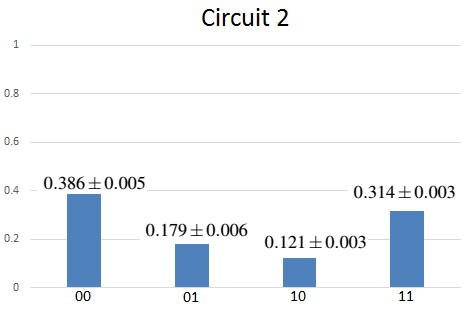}
\end{center}

\begin{center}
	\includegraphics[scale=0.55]{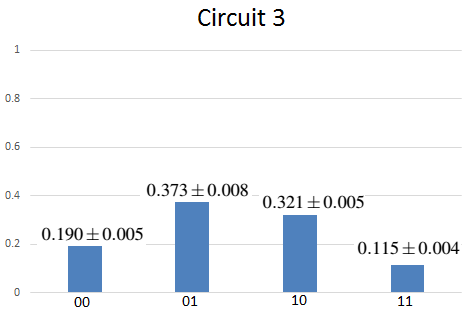} 
	
	\justifying\noindent\begin{small}\textbf{Figure 12}. Mean probability outcomes ($\pm$ standard deviation) of circuits 1,2,3 from Fig.11  after 5 runs of 8192 shots on the 5-qubit IBM computer (ibmqx4), using qubits 2 and 1.\end{small}
\end{center}

\vspace{0.1cm}

Therefore, the experimental values found for $P(\vec{a},\vec{b})$, $P(\vec{a},\vec{c})$ and $P(\vec{b},\vec{c})$ are \footnote{The errors have been calculated assuming the statistical independence of the variables summed: therefore, the variances have been added and then the square root was taken to obtain the standard deviation of the sum.}

\beq \label{b1} \begin{split} &P(\vec{a},\vec{b})_{exp}= -0.392\pm0.014 \quad,\\ \\  &P(\vec{a},\vec{c})_{exp}=0.401\pm0.009 \quad,\\ \\ &P(\vec{b},\vec{c})_{exp}=-0.389 \pm0.012 \quad,\end{split}\eeq

\noindent where $P(\vec{a},\vec{b})$ is the sum of the probabilities of finding the results $00$ or $11$ minus the probabilities of finding $01$ or $10$, and similarly for $P(\vec{a},\vec{c})$ and $P(\vec{b},\vec{c})$ (note that the results according to Quantum Mechanics should be $\pm0.5$). Hence, Bell's inequality is violated:

\beq \label{b2} 1.182\pm0.020 \nleq1 \quad\; .\eeq 

\vspace{-0.4cm}

Other Bell-type inequalities can be tested on the IBM 5Q as well; for instance, the CHSH inequality \cite{CHSH,CHSHQ}.

\vspace{0.7cm}

\noindent\textbf{4. Mermin's inequalities}

\vspace{0.1cm}

Mermin's inequalities are a generalization of Bell-type inequalities for systems of more than two spin 1/2 particles (or qubits), derived in 1990 by Mermin \cite{MerminI}. The basic idea is that there exists an operator $M_n$ (called a Mermin polynomial) whose quantum-mechanical expectation value exceeds the limits imposed by Local Realism for certain states. In this way, the incompatible predictions of the two theories can be experimentally tested and confronted: the inequalities must be fulfilled if Local Realism holds and violated in case Quantum Mechanics is correct.

The quantum-mechanical expectation value for $M_n$ for some highly entangled states (GHZ-type states) exceeds the bounds imposed by Local Realism by an amount that grows exponentially with $n$ (where $n$ is the number of qubits). This fact implies that there is (in principle) no apparent limit to the amount by which Bell-type inequalities can be violated by certain entangled states. However, because the joint efficiency of $n$ measurement apparatuses necessarily declines exponentially in $n$, and the $n$-qubit GHZ states are increasingly difficult to prepare as $n$ grows, an exponentially greater violation of the inequalities for higher values of $n$ will hardly be observed \cite{MerminI}. Mermin's inequalities have been proposed as a figure of merit to assess the fidelity of a quantum computer \cite{Latorre}; in fact, they can be tested on the IBM 5Q computers  for $n=3,4,5\,$.


For $n=3$, a GHZ-type state and associated Mermin polynomial that give maximal violation of the corresponding inequality are:

\beq \label{m3} \begin{split} &|\psi\ra=\frac{1}{\sqrt{2}}\,(|000\ra+i\,|111\ra)\quad, \\\\ M_3=\sigma_y^2\sigma_x^1&\sigma_x^0+\sigma_x^2\sigma_y^1\sigma_x^0+\sigma_x^2\sigma_x^1\sigma_y^0-\sigma_y^2\sigma_y^1\sigma_y^0\quad,\\\\\\ &\quad\quad\quad\;\la M_3\ra\leq 2\quad.\end{split} \eeq

For $n=4$, those are \cite{Latorre}:

\beq \label{m4} \begin{split} |\psi\ra&=\frac{1}{\sqrt{2}}\,(|0000\ra+e^{\,i\pi/4}\,|1111\ra)\quad,\\\\ M_4= (\sigma_y^3\sigma_x^2\sigma_x^1\sigma_x^0+&\sigma_x^3\sigma_y^2\sigma_x^1\sigma_x^0+\sigma_x^3\sigma_x^2\sigma_y^1\sigma_x^0+\sigma_x^3\sigma_x^2\sigma_x^1\sigma_y^0)+\\\\+(\sigma_y^3\sigma_y^2\sigma_x^1\sigma_x^0&+\sigma_y^3\sigma_x^2\sigma_y^1\sigma_x^0+\sigma_y^3\sigma_x^2\sigma_x^1\sigma_y^0+\sigma_x^3\sigma_y^2\sigma_y^1\sigma_x^0+\\\\+\sigma_x^3\sigma_y^2\sigma_x^1\sigma_y^0&+\sigma_x^3\sigma_x^2\sigma_y^1\sigma_y^0)-\sigma_y^3\sigma_y^2\sigma_y^1\sigma_y^0-\sigma_x^3\sigma_x^2\sigma_x^1\sigma_x^0-\\\\-(\sigma_y^3\sigma_y^2\sigma_y^1\sigma_x^0&+\sigma_y^3\sigma_y^2\sigma_x^1\sigma_y^0+\sigma_x^3\sigma_y^2\sigma_y^1\sigma_y^0+\sigma_x^3\sigma_y^2\sigma_y^1\sigma_y^0)\;\;,\\\\\\ &\quad\quad\quad\;\;\la M_4\ra\leq4\quad.\end{split} \eeq

And for $n=5$:

\beq \label{m5} \begin{split} |\psi\ra=&\frac{1}{\sqrt{2}}\,(|00000\ra+i\,|11111\ra) \quad,\\\\ M_5=(\sigma_y^4\sigma_x^3\sigma_x^2\sigma_x^1\sigma_x^0&+\sigma_x^4\sigma_y^3\sigma_x^2\sigma_x^1\sigma_x^0+\sigma_x^4\sigma_x^3\sigma_y^2\sigma_x^1\sigma_x^0+\\\\+\sigma_x^4\sigma_x^3\sigma_x^2\sigma_y^1\sigma_x^0+&\sigma_x^4\sigma_x^3\sigma_x^2\sigma_x^1\sigma_y^0)-(\sigma_y^4\sigma_y^3\sigma_y^2\sigma_x^1\sigma_x^0+\\\\+\sigma_y^4\sigma_y^3\sigma_x^2\sigma_y^1\sigma_x^0&+\sigma_y^4\sigma_y^3\sigma_x^2\sigma_x^1\sigma_y^0+\sigma_y^4\sigma_x^3\sigma_y^2\sigma_y^1\sigma_x^0+\\\\+\sigma_y^4\sigma_x^3\sigma_y^2\sigma_x^1\sigma_y^0&+\sigma_y^4\sigma_x^3\sigma_x^2\sigma_y^1\sigma_y^0+\sigma_x^4\sigma_y^3\sigma_y^2\sigma_y^1\sigma_x^0+\\\\+\sigma_x^4\sigma_y^3\sigma_y^2\sigma_x^1\sigma_y^0&+\sigma_x^4\sigma_y^3\sigma_x^2\sigma_y^1\sigma_y^0+\sigma_x^4\sigma_x^3\sigma_y^2\sigma_y^1\sigma_y^0)+\\\\&+\sigma_y^4\sigma_y^3\sigma_y^2\sigma_y^1\sigma_y^0\quad,\\\\\\&\quad\quad\la M_5\ra\leq4\quad.\end{split}\eeq

\vspace{0.2cm}
The circuits required for testing Mermin's inequality for $n=3$ and the choice of settings (\ref{m3}) are shown in Fig.13 (circuits for $n=4$ and $n=5$ are analogous):

\begin{center}
	\includegraphics[scale=0.37]{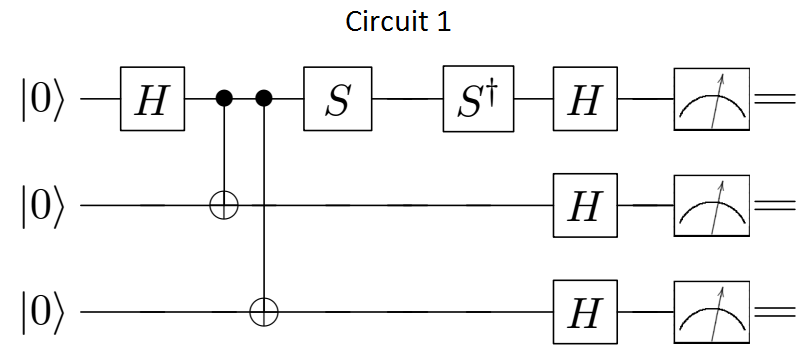}
	\includegraphics[scale=0.37]{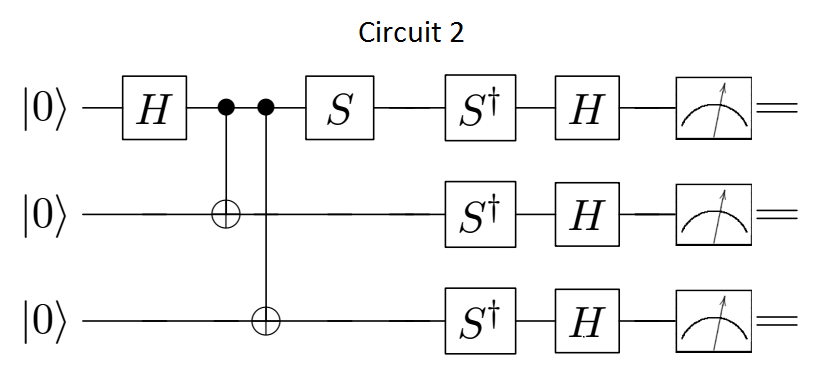}
	
	\justifying\noindent\begin{small}\textbf{Figure 13}. Circuits for testing Mermin's inequality for $n=3$: Circuit 1 measures $\sigma_y^2\sigma_x^1\sigma_x^0$ ($YXX$) and Circuit 2 measures $\sigma_y^2\sigma_y^1\sigma_y^0$ ($YYY$).\end{small}
\end{center}

The Hadamard gate, cNOTs and $S$ gate in Fig.13 generate the state $|\psi\ra=\frac{1}{\sqrt{2}}\,(|000\ra+i\,|111\ra)$. Changing the position of the $S^\dagger$ gates in Circuit 1 from Fig.13 allows to measure $\sigma_x^2\sigma_y^1\sigma_x^0$ and $\sigma_x^2\sigma_x^1\sigma_y^0$. However, from our own results and those obtained in \cite{Latorre} it seems safe to assume symmetry under qubit exchange. That is: $\la\sigma_y^2\sigma_x^1\sigma_x^0\ra_{exp}\simeq\la\sigma_x^2\sigma_y^1\sigma_x^0\ra_{exp}\simeq\la\sigma_x^2\sigma_x^1\sigma_y^0\ra_{exp}$, and only one circuit is needed to measure the three terms in the polynomial. The same is true for $n=4$ and $n=5$, which makes it possible to use a single circuit to measure all the terms with a same number of $\sigma_y$s.

The experimental results obtained for the states and polynomials (\ref{m3}), (\ref{m4}) and (\ref{m5}) on the ibmqx2 are collected in Table 3. These results show a clear violation of Mermin's inequalities for $n=3,4,5$. They can be compared to those found by Alsina and Latorre \cite{Latorre}, also shown in Table 3 (the choice of state and polynomial for $n=5$ is slightly different in \cite{Latorre}, but completely equivalent to (\ref{m5}), since it has the same quantum-mechanical expectation value and Local Realism bound). We interpret the better results presented here to reflect the improvements of the IBM chips during the months in between both publications \footnote{In particular, we believe that the improvements in the violation of Mermin's inequalities are due to the availability of cNOTs among qubits 1,0 and 3,4, which were not present at the time Alsina and Latorre obtained their results. This interpretation is consistent with the fact that gate errors, readout errors, and decoherence and relaxation times were similar in the present work and in \cite{Latorre}, and with the fact that no improvement has been obtained for the 3-qubit case}.






\vspace{0.1cm}
\begin{table}[H]
	\begin{center}
		\begin{tabular}{c|c|c|c|c|}
			       &\, LR\, & \,QM \,& \quad A-L\quad\quad & \quad\quad GM-S \quad\quad \\ \hline
			3 qubits\, & 2  & 4  & \;$2.85\pm0.02$\; & $2.84\pm0.07$ \\ \hline
			4 qubits\, & 4 & 8$\sqrt{2}$ &\;$4.81\pm0.06$\; & $5.42\pm0.04$ \\ \hline
			5 qubits \,& 4 & 16 & \;$4.05\pm0.06$ \;& $7.06\pm0.03$ \\ \hline
		\end{tabular}
		\end{center}
		\begin{small}\textbf{Table 3}. Comparison with the results obtained in \cite{Latorre} for Mermin's inequalities. LR stands for Local Realism bound, QM for Quantum Mechanics bound, A-L for experimental results obtained by Alsina and Latorre and GM-S for experimental results obtained in the present work.\end{small}
\end{table}

\noindent\textbf{5. Prime state}

\vspace{0.1cm}


The prime state $|p_n\ra$ is the superposition of all the computational-basis states that correspond to prime numbers (written in binary format) up to a certain value $N=2^n-1$ \cite{Primes}:

\beq |p_n\ra =\frac{1}{\sqrt{\pi(2^n)}}\,\sum_{x\in prime}^{2^n-1} |x\ra \quad\;,\eeq

\noindent where $\pi(x)$, known as the prime counting function, is the number of primes less or equal than $x$. This state bears a large amount of entanglement \cite{Eprimes}. The prime state for $n=3$, i.e. $|p_3\ra=\frac{1}{2}\,(|2\ra+|3\ra+|5\ra+|7\ra)=\frac{1}{2}\,(|010\ra+|011\ra+|101\ra++|111\ra)$, can be created with the following circuit:

\begin{center}
	\includegraphics[scale=0.4]{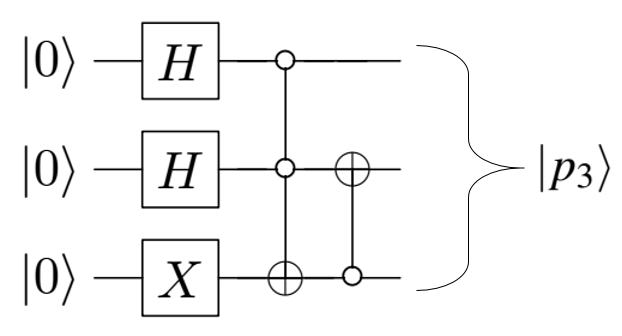}
	
	\justifying\noindent\begin{small}\textbf{Figure 14}. Circuit that creates the prime state $|p_3\ra$ on the three uppermost qubits using an ancilla.\end{small}
\end{center}

The $X$ gate acting on the first qubit ensures that all number states in the superposition created with the two Hadamard gates are odd: after these three gates have acted, the system is in the state $|\psi\ra=\frac{1}{2}\,(|001\ra+|011\ra+|101\ra+|111\ra)$. Thus, all that remains is converting the term $|001\ra$ (1 is not a prime number) into $|010\ra$. This is achieved with the help of a Toffoli gate and a cNOT.

After implementing the circuit of Fig.14 for creating the prime state $|p_3\ra$ on the ibmqx4, a joint measurement of the three qubits in the computational basis was carried out in order to assess the overall performance of the circuit. The results are shown in Fig.15.

\begin{center}
	\includegraphics[scale=0.65]{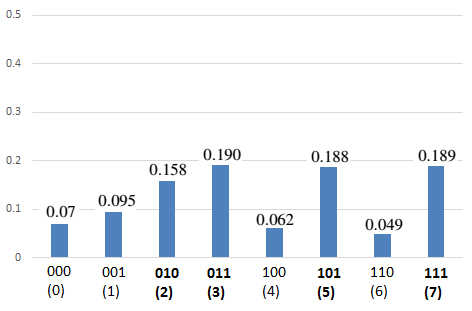}
	
	\justifying\noindent\begin{small}\textbf{Figure 15}. Mean probability outcomes of joint measurements in the computational basis of the three qubits of the purported prime state $|p_3\ra$ after 5 runs of 8192 shots on the 5-qubit IBM quantum computer (ibmqx4), using qubits 2,1,0 (standard deviations are not shown for the sake of clarity). In bold are the outcomes corresponding to prime numbers.\end{small}
\end{center}


The results show that, upon measurement of the purported prime state $|p_3\ra$, a prime number is obtained with probability $72.5\pm0.7 \%$ \footnote{The error has been calculated as in Bell's inequality}. In this sense it can be said that the prime state $|p_3\ra$ has been constructed with an approximated accuracy of $73\%$.

The interest in constructing prime states (something that can in principle be done efficiently using Grover's algorithm with a primality test as oracle) is that they would allow to experimentally test, for instance, Riemann hypothesis --one of the mathematical problems of the millennium \cite{Eprimes}. But in order to falsify Riemann hypothesis (or extend the limits of its validity) it is necessary to build superposition states of billions of prime numbers. Constructing the state $|p_3\ra$, even if this is not done by a general method for creating prime states, thus seems a modest first step.

Moreover, there exist a number of quantities in Number Theory that can surprisingly be measured experimentally on prime states \cite{Primes}. The mean value of $\sigma_z^1$ is:

\beq \la\sigma^1_z\ra=\frac{\pi_{4,1}(N)-\pi_{4,3}(N)-1}{\pi(N)}\quad,\eeq 

\noindent where $\pi_{4,1}(N)$ is the number of primes less or equal than $N$ that can be written as $4m+1$ with $m$ a positive integer (i.e. that are equal to 1 (mod 4)) and $\pi_{4,3}(N)$ is the number of primes less or equal than $N$ that can be written as $4m+3$ with $m$ a non-negative integer (i.e. that are equal to 3 (mod 4)). Thus $\pi(N)=\pi_{4,1}(N)+\pi_{4,3}(N)+1$. The difference between these two quantities $\Delta(N)\equiv\pi_{4,3}(N)-\pi_{4,1}(N)$ is known as the Chebyshev bias, and it is curiously positive for most values of $N$ for which it has been calculated so far \cite{Chebyshev} (the name is due to Pafnuty Chebyshev, who first noticed that the remainder upon dividing the primes by 4 gives 3 more often than 1).

Therefore, measuring the second qubit in the computational basis, and taking an outcome 0 as a 1 and an outcome 1 as a $-1$ when computing the mean value, give the Chebyshev bias, provided that $\pi(N)$ is known. For the state $|p_3\ra$ created on the ibmqx4, the experimental result for $\la\sigma^1_z\ra$ obtained is:

\beq \la\sigma^1_z\ra_{exp}= -0.301\pm0.007\quad,\eeq

\noindent while the theoretical expected value for $\la\sigma^1_z\ra$ is:

\beq  \la\sigma^1_z\ra_{th}=-0.500 \quad,\eeq

\noindent which gives a relative error of around $40\%$ in the measurement. Other quantities that can be measured experimentally on the prime state are:

\beq \label{p} \la\sigma^1_x\ra=\frac{2\,\pi_2^{(1)}(N)}{\pi(N)}\;\;\;\;\;,\;\;\;\;\;\la\sigma_x^1\sigma_x^2+\sigma_y^1\sigma_y^2\ra=\frac{4\,\pi_2^{(3)}(N)}{\pi(N)}\;\;\;\,, \eeq

\noindent where $\pi_2^{(1)}(N)$ is the number of twin prime pairs $(p,\,p+2)$ less or equal than $N$ with $p=1$ (mod 4) and $\pi_2^{(3)}(N)$ is the number of twin prime pairs $(p,\,p+2)$ less or equal than $N$ with $p=3$ (mod 4). The sum $\pi_2^{(1)}(N)+\pi_2^{(3)}(N)$ is equal to the number of twin prime pairs $\pi_2(N)$. In analogy with the Chebyshev bias the twin prime bias is defined as $\Delta_2(N)=\pi_2^{(3)}(N)-\pi_2^{(1)}(N)$.

The experimental results of the measurements of operators (\ref{p}) on $|p_3\ra$ on the ibmqx4 are:

\beq \begin{split}\la\sigma^1_x\ra_{exp}=0.435\pm0.006 \quad,\\ \\ \la\sigma_x^1\sigma_x^2+\sigma_y^1\sigma_y^2\ra_{exp}= 0.641\pm0.022\quad, \end{split}\eeq

\noindent while the theoretical expected values are:

\beq \la\sigma^1_x\ra_{th}=0.500\;\;\;\;\;,\;\;\;\;\;\la\sigma_x^1\sigma_x^2+\sigma_y^1\sigma_y^2\ra_{th}=1.000 \;\;\;\,,\eeq

\noindent which give relative errors of approximately $13\%$ and $36\%$ respectively.


\section{Conclusions}

The time of Quantum Computation has come. Quantum-computer prototypes have already been constructed and some of them are even available on the cloud, thanks to the IBM Quantum Experience. This allows to perform experiments and assess the functioning of these first quantum computers. In the present work, the protocol of dense coding was completed using superconducting qubits for the first time, with efficiencies around $74\%$ in the worst case. Quantum Fourier transforms have also been implemented; although QFTs were used in \cite{Benchmark, permutation, Cloud}, they were not performed on states of more than two qubits due to previous limitations of the IBM Quantum Experience. Moreover, the performance of QFTs on the IBM 5Q has never been assessed explicitly before. The original Bell's inequality and Mermin's inequalities up to $n=5$ were checked and shown to violate Local Realism, with an improvement with respect to the results found in \cite{Latorre} that we interpret as a reflection of the improvements of the IBM quantum computers during the last months. Finally, the construction of the prime state $|p_3\ra$ have been carried out, which constitutes the first experimental realization of a prime state. Overall, the results obtained in these experiments, although moderately good in most cases, are still far from optimum. 

Therefore, in light of these results, it is clear that there is still a lot of work to be done before a quantum computer can actually be useful for solving mathematical problems, simulating efficiently quantum systems, breaking classical encryption systems, etc (in other words, fully achieve so-called quantum supremacy \cite{Preskill}). But given the astounding pace at which technological developments are being push forward, it seems that the dream of building a functional universal quantum computer within the next twenty years is close. Even more when one takes into account that it is likely (or at least plausible) that the full power of IBM quantum computers has not been shown yet for commercial reasons and thus it is not exhibited by the chips of the IBM Quantum Experience.

With no known fundamental obstacles on the way, quantum computers will surely end up being a reality in research centres all around the world. And, as it happens every time a new regime of Nature becomes experimentally available, a plethora of new discoveries will certainly accompany this `Second Quantum Revolution'. Meanwhile, proofs of principle like the ones presented here for several quantum circuits will be useful to help improving the systems.

\section{Acknowledgements}

We thank J.I. Latorre and D. Alsina for useful conversations. We also thank the IBM Quantum Experience for the use of the ibmqx2 and ibmqx4. G.S. acknowledges the  grants FIS2015-69167-C2-1-P from the Spanish government, QUITEMAD+ S2013/ICE-2801 from the Madrid regional government and SEV-2016-0597 of the Centro de Excelencia Severo Ochoa Programme.


\end{document}